\documentstyle[12pt]{article}
\begin{document}
\title{\bf Drag Force of Moving Quark in STU Background}
\author{J. Sadeghi $^{a,}$\thanks{Email:
pouriya@ipm.ir}\hspace{1mm}, M. R. Setare $^{b,}$\thanks{Email:
rezakord@ipm.ir}\hspace{1mm}, B. Pourhassan $^{a}$\thanks{Email:
b.pourhassan@umz.ac.ir}\hspace{1mm} and
S. Hashmatian $^{a}$\thanks{Email: s.heshmatian@umz.ac.ir}\\
$^a$ {\small {\em  Sciences Faculty, Department of Physics, Mazandaran University,}}\\
{\small {\em P .O .Box 47415-416, Babolsar, Iran}}\\
$^b$ {\small {\em Department of Science, Payame Noor University,
Bijar, Iran,}}} \maketitle
\begin{abstract}
\noindent In this paper we consider a quark moving in $D=5$,
${\mathcal{N}}=2$ supergravity thermal plasma. By using three charge
non-extremal black hole solution (STU solution) we calculate drag
force on the quark and diffusion constant from $AdS$/CFT
correspondence.
\end{abstract}
\section{Introduction}
$AdS$/CFT correspondence, which is one of the interesting dualities
in physics, was first suggested by Maldacena [1] and then developed
by Witten and Gubser et al. [2, 3]. The most famous case of
$AdS$/CFT correspondence is the duality between IIB string theory in
$AdS_{5}$ $\times S^{5}$ space and ${\mathcal{N}}=4$ supersymmetric
Yang -
Mills gauge theory on $4$ dimensional boundary of $AdS_{5}$ space [4 - 6].\\
One of the various applications of $AdS$/CFT correspondence is in
QCD. So that by using string theory description one can attack the
complicated problem of QCD. One of these complicated problems is
moving charged particles in medium. If one would like to consider a
moving quark in plasma, there are some problems in QCD description,
while for instance the energy loose of a heavy quark moving in
supersymmetric Yang - Mills thermal plasma has been investigated
easily from $AdS$/CFT correspondence [7-13]. This problem is also
considered in the case of a quark - antiquark pair [7, 14 - 16] and
the jet quenching parameter has been calculated [17 - 24]. According
to Maldacena's dictionary we have the following model; a moving
quark in thermal plasma is considered such as a string in $AdS$
space stretched from $D$-brane to the black hole horizon. The end
point of string attached to $D$-brane represents a quark. Drag force
on the quark due to plasma is interpreted as a current of energy and
momentum along the string. The energy and momentum is
transformed from the string end point to the horizon.\\
Correspondence between ${\mathcal{N}}=2$ supergravity theory and
string theory will be interesting in physics [25, 26]. Actually,
solutions of ${\mathcal{N}}=2$ supergravity may be solutions of
supergravity theory with more supersymmetry i.e. ${\mathcal{N}}=4$
and ${\mathcal{N}}=8$. The ${\mathcal{N}}=2$ supergravity theory in
five dimensions can be obtained by compactifying eleven dimensional
supergravity in a three-fold Calabi-Yau [27]. Also Anti de Sitter
supergravity can be obtained by gauging the $U(1)$ subgroup of the
$SU(2)$ group in
${\mathcal{N}}=2$ supersymmetric algebra.\\
Here we consider a moving quark in $D=5$, ${\mathcal{N}}=2$
supergravity thermal plasma and suppose that the quark is influenced
by an external force $F$. It has the momentum $P$, mass $M$ and
velocity $v$, also the friction coefficient of plasma is $\mu$. We
can write the equation of motion in the form of $\dot{p}=F-\mu
p=F-\mu Mv$. To obtain some information about drag force, it is
useful to consider two special cases. One, we assume that momentum
of the charged particle is constant. Hence by using $F=\mu p$ for
particle with mass $M$ and non-relativistic momentum $p=Mv$ one can
obtain $\mu M=\frac{F}{v}$. So, by measuring the velocity of
particle for a given force we can obtain $\mu M$. It shows that we
can't find $\mu$ independently. Second, we assume that external
force does not exist, so from equation of motion one can find
$p(t)=p(0)e^{-\mu t}$. In another word by measuring the ratio
$\frac{\dot{p}}{p}$ or $\frac{\dot{v}}{v}$ we can determine $\mu$
without any dependance to $M$. These lead us to obtain drag force
for a moving quark in plasma [7, 8].\\
In order to calculate drag force by using $AdS$/CFT correspondence
we will consider three charges non-extremal black hole solution (STU
model) [26, 28]. STU solution obtained by Behrndt et al. as special
solution of equation of motion
for $D=5$, ${\mathcal{N}}=2$ gauge supergravity theory [26].\\
This paper is organized as follows, in section 2 we introduce STU
model and obtain canonical momentum densities for non-extremal black
hole background. Then in section 3 we will obtain drag force and
diffusion constant for various states of string corresponding to the
quark. Quasi-normal modes of string considered in section 4, and
finally in section 5 we have conclusion and discussion,  also we
give some suggestions for future works.
\section{STU Model Solution}
STU model solution is given by the following background metric [28],
\begin{eqnarray}\label{s1}
ds^{2}&=&-\frac{f_{k}}{{\mathcal{H}}^{\frac{2}{3}}}dt^{2}+{\mathcal{H}}^{\frac{1}{3}}(\frac{dr^{2}}{f_{k}}+r^{2}d\Omega_{3,k}^{2}),\nonumber\\
f_{k}&=&k-\frac{m_{k}}{r^{2}}+\frac{r^{2}}{L^{2}}{\mathcal{H}},
\end{eqnarray}
where ${\mathcal{H}}=H_{1}H_{2}H_{3}$ and
$H_{i}=1+\frac{q_{i}}{r^{2}}$ ($i=1, 2, 3$). Here we have three real
scalar field $X^{i}=\frac{{\mathcal{H}}^{\frac{1}{3}}}{H_{i}}$ which
satisfy $X^{1}X^{2}X^{3}=1$. $r$ is a radial coordinate along the
black hole and $f_{k}(r)=0$ shows that the horizon of black hole is
in $r=r_{h}$.\\
Three charges $q_{i}$ are related to three independent angular
momentum in ten dimensions and $k$ indicates the curvature of space.
$k=1$ shows a three dimensional sphere with unit radius and $k=0$
indicates three dimensional flat real space. Using Kaluza-Klein
dimension reduction of near-extremal spinning brane, we obtain the
corresponding solution of this space [29]. Finally for $k=-1$ we
have a metric of a pseudo sphere, ie the space is in hyperbolic form.\\
At first, if we consider $k=1$ and suppose three charges to be equal
( $q_{1}=q_{2}=q_{3}=q$) the results would be similar to Ref. [30,
31] and in this case $X^{1}=X^{2}=X^{3}=1$. In this paper we will
try to solve the general case with $k$. From Ref. [26] we know that
$d\Omega_{3,k}^{2}=d\chi^{2}+(\frac{\sin{\sqrt{k}\chi}}{\sqrt{k}})^{2}(d\theta^{2}+\sin^{2}{\theta}d\phi^{2})$,
we would like to assume that the motion in manifold with metric
$d\Omega_{3,k}^{2}$ is only in a transverse axis which we called
$x$. Hence in this special case one can write
$d\Omega_{3,k}^{2}\rightarrow dx^{2}$.\\
Also we assume that only one charge exists, $q_{1}=q$
($q_{2}=q_{3}=0$), then we can write,
\begin{eqnarray}\label{s2}
ds^{2}&=&-\frac{f_{k}}{H^{\frac{2}{3}}}dt^{2}+H^{\frac{1}{3}}(\frac{dr^{2}}{f_{k}}+r^{2}dx^{2}),\nonumber\\
f_{k}&=&k-\frac{m_{k}}{r^{2}}+\frac{r^{2}}{L^{2}}H,\nonumber\\
H&=&1+\frac{q}{r^{2}}.
\end{eqnarray}
As we know the dynamics of an open string is described by Nambo-Goto
action,
\begin{equation}\label{s3}
S=-T_{0}\int{d\tau d\sigma \sqrt{-g}},
\end{equation}
where $T_{0}$ is the string tension and $\sigma$, $\tau$ are
coordinates of string world sheet. Also $g_{ab}$ is metric on world sheet of string, and $g=det{g_{ab}}$.\\
Using the static gauge we set $\tau=t$, $\sigma=r$,then string world
sheet is described by $x(t,r)$, so one can obtain,
\begin{equation}\label{s4}
-g=\frac{1}{H^{\frac{1}{3}}}-\frac{H^{{\frac{2}{3}}}
r^{2}}{f_{k}L^{2}}{\dot{x}}^{2}+\frac{f_{k}
r^{2}}{L^{2}H^{\frac{1}{3}}}{x^{\prime}}^{2},
\end{equation}
and lagrangian density is ${\mathcal{L}}=-T_{0}\sqrt{-g}$. Therefore
the string equation of motion is as following,
\begin{equation}\label{s5}
\frac{\partial}{\partial r}(\frac{f_{k}
r^{2}}{L^{2}H^{\frac{1}{3}}\sqrt{-g}}x^{\prime})-\frac{H^{\frac{2}{3}}r^{2}}{f_{k}L^{2}}\frac{\partial}{\partial
t}(\frac{\dot{x}}{\sqrt{-g}})=0.
\end{equation}
But in order to obtain total energy and momentum of string and drag
force we have to calculate the canonical momentum densities,
\begin{eqnarray}\label{s6}
{\pi}_{t}^{0} &=& -\frac{T_{0}}{\sqrt{-g}}\frac{(1+\frac{f_{k}r^{2}}{L^{2}}{x^{\prime}}^{2})}{H^{\frac{1}{3}}},\nonumber\\
{\pi}_{x}^{0} &=& \frac{T_{0}}{\sqrt{-g}}\frac{H^{\frac{2}{3}}r^{2}}{f_{k}L^{2}}\dot{x},\nonumber\\
{\pi}_{r}^{0} &=&
-\frac{T_{0}}{\sqrt{-g}}\frac{H^{\frac{2}{3}}r^{2}}{f_{k}L^{2}}\dot{x}x^{\prime},
\end{eqnarray}
and
\begin{eqnarray}\label{s7}
{\pi}_{t}^{1} &=& \frac{T_{0}}{\sqrt{-g}}\frac{f_{k}r^{2}}{L^{2}H^{\frac{1}{3}}}\dot{x}x^{\prime},\nonumber\\
{\pi}_{x}^{1} &=& -\frac{T_{0}}{\sqrt{-g}}\frac{f_{k}r^{2}}{L^{2}H^{\frac{1}{3}}}x^{\prime},\nonumber\\
{\pi}_{r}^{1} &=&
\frac{T_{0}}{\sqrt{-g}}(-\frac{1}{H^{\frac{1}{3}}}+\frac{H^{\frac{2}{3}}r^{2}}{f_{k}L^{2}}{\dot{x}}^{2}).
\end{eqnarray}
Then we can find drag force by $\dot{p}=\pi_{x}^{1}$. Also the total
energy and momentum of string are described by following relations,
\begin{eqnarray}\label{s8}
E&=&-\int{dr{\pi}_{t}^{0}},\nonumber\\
p&=&\int{dr{\pi}_{x}^{0}}.
\end{eqnarray}
In the two next sections we use the above relations for some special
cases and will obtain drag force, diffusion constant, total energy
and momentum of string.
\section{Straight and Curved String}
In this section we will consider three different cases for a string
and calculate drag force. As we mentioned, a moving quark in
${\mathcal{N}}=2$ supergravity thermal plasma is correspondent to
the string which stretches from $D$-brane to black hole, so the end
point of string on $D$-brane represents the quark. It means that
according to Maldacena dictionary, a quark in CFT is equal to a
string in $AdS$ space. In the other hand existence of quark flavor
causes to have a $D$-brane, and by adding temperature to plasma in
CFT we have a black hole in $AdS$ space.\\
At first we consider a static quark in plasma which is equal to a
static string stretched directly from $r=r_{m}$ on $D$-brane to
$r=r_{h}$ in black hole. It means that we deal with the simplest
solution of equation (5), $x(t,r)=x_{0}$ where $x_{0}$ is a
constant. In this case total momentum of string and drag force are
zero and since $-g=H^{-\frac{1}{3}}$ we have total energy of string,
by using relations (6), (7) and (8),
\begin{equation}\label{s9}
E=T_{0}\left[\frac{3r(1+\frac{r^{2}}{q})^{\frac{1}{6}}}{4(1+\frac{q}{r^{2}})^{\frac{1}{6}}}
2F_{1}(\frac{2}{3}, \frac{1}{6}, \frac{5}{3},
-\frac{r^{2}}{q})\right]_{r_{h}}^{r_{m}},
\end{equation}
where $2F_{1}(\frac{2}{3}, \frac{1}{6}, \frac{5}{3},
-\frac{r^{2}}{q})$ is a hypergeometric function. It should be
mentioned that in limit of $q\rightarrow r$ one can obtain
$E=T_{0}(r_{m}-r_{h})$, which agrees with the case of a moving quark
in ${\mathcal{N}}=4$ SYM thermal plasma [7, 31].\\
In the second case the straight string is moving with constant
velocity, $x(r, t)=x_{0}+vt$ which can be the solution of equation
(5), so we have,
\begin{equation}\label{s10}
-g=\frac{1}{H^{\frac{1}{3}}}-\frac{H^{\frac{2}{3}}
r^{2}}{f_{k}L^{2}}v^{2}.
\end{equation}
Then canonical momentum densities are,
\begin{eqnarray}\label{s11}
{\pi}_{t}^{0} &=& -\frac{T_{0}}{\sqrt{-g}}\frac{1}{H^{\frac{1}{3}}},\nonumber\\
{\pi}_{x}^{0} &=& \frac{T_{0}}{\sqrt{-g}}\frac{H^{\frac{2}{3}}r^{2}v}{f_{k}L^{2}},\nonumber\\
{\pi}_{r}^{1} &=& \frac{T_{0}}{\sqrt{-g}}(\frac{H^{\frac{2}{3}}
r^{2}}{f_{k}L^{2}}v^{2}-\frac{1}{H^{\frac{1}{3}}}),
\end{eqnarray}
and $\pi_{r}^{0}=\pi_{t}^{1}=\pi_{x}^{1}=0$. However in that case we
have imaginary $\sqrt{-g}$, so it is not
physical solution [7, 31], therefore we consider the other case.\\
The most real and important case is a curved string moving with
constant velocity $v$. So the following relation,
\begin{equation}\label{s12}
x(r,t) =x(r)+vt,
\end{equation}
satisfies equation (5) and leads us to have following expression,
\begin{equation}\label{s13}
-g=\frac{1}{H^{\frac{1}{3}}}-\frac{H^{{\frac{2}{3}}}
r^{2}}{f_{k}L^{2}}v^{2}+\frac{f_{k}
r^{2}}{L^{2}H^{\frac{1}{3}}}{x^{\prime}}^{2},
\end{equation}
therefore from equation (5) we have,
\begin{equation}\label{s14}
\frac{\partial}{\partial r}(\frac{f_{k}
r^{2}}{H^{\frac{1}{3}}L^{2}v\sqrt{-g}}x^{\prime})=0.
\end{equation}
The solution of equation (14) can be obtain by,
\begin{equation}\label{s15}
{x^{\prime}}^{2}=\frac{L^{2}C^{2}v^{2}}{f_{k}^{2}
r^{4}}\frac{f_{k}L^{2}H^{\frac{1}{3}}-H^{\frac{4}{3}}r^{2}v^{2}}{f_{k}-\frac{L^{2}C^{2}v^{2}H^{\frac{1}{3}}}{r^{2}}},
\end{equation}
where $C$ is constant, and it's easy to obtain,
\begin{eqnarray}\label{s16}
{\pi}_{t}^{1} &=& \frac{T_{0}Cv^{2}}{L^{2}},\nonumber\\
{\pi}_{x}^{1} &=& -\frac{T_{0}Cv}{L^{2}}.
\end{eqnarray}
Now the important problem is to calculate constant $C$. It can be
obtained by the condition $-g\geq 0$, otherwise we have imaginary
action, energy and momentum. By using ${x^{\prime}}^{2}$ in equation
(13) in case of $-g> 0$, we find $C$ as follows,
\begin{equation}\label{s17}
C=\pm(\frac{r_{c}}{L})^{2}(1+\frac{q}{r_{c}^{2}})^{\frac{1}{3}},
\end{equation}
where ${r_{c}}$ is root of equation,
\begin{equation}\label{s18}
r^{4}(1-v^{2})(1+\frac{q}{r^{2}})+kL^{2}r^{2}-m_{k}L^{2}=0,
\end{equation}
which is minimum radius ($-g$ to be zero). In order to obtain
${\pi}_{t}^{1}$ and ${\pi}_{x}^{1}$ we use equations (16) and (17),
\begin{eqnarray}\label{s19}
{\pi}_{t}^{1} =
T_{0}\frac{v^{2}}{L^{4}}r_{c}^{2}(1+\frac{q}{r_{c}^{2}})^{\frac{1}{3}},\nonumber\\
{\pi}_{x}^{1} =
-T_{0}\frac{v}{L^{4}}r_{c}^{2}(1+\frac{q}{r_{c}^{2}})^{\frac{1}{3}},
\end{eqnarray}
which leads us to have following equation,
\begin{equation}\label{s20}
\dot{p}={\pi}_{x}^{1} =-\mu p=-\mu Mv.
\end{equation}
In other word drag force coefficient is given by,
\begin{equation}\label{s21} \mu
M=\frac{T_{0}}{L^{4}}r_{c}^{2}(1+\frac{q}{r_{c}^{2}})^{\frac{1}{3}},
\end{equation}
and we can also calculate the quark diffusion constant
$D=\frac{T}{\mu M}$ where $T$ is Hawking temperature of black hole.
So we have,
\begin{equation}\label{s22}
D=\frac{TL^{4}}{T_{0}}\frac{1}{r_{c}^{2}(1+\frac{q}{r_{c}^{2}})^{\frac{1}{3}}}.
\end{equation}
Next we are going to consider special condition,\\
we take $q\rightarrow 0$ limit in which there is a relation between
$ {\mathcal{N}}=2$ supergravity and ${\mathcal{N}}=4$ super Yang-
Mills theory [32]. In this limit $H\longrightarrow 1$ and we have,
\begin{equation}\label{s23}
r_{c}^{2}=\frac{L}{2(1-v^{2})}\left(-kL\pm
\sqrt{k^{2}L^{2}+4m_{k}(1-v^{2})}\right),
\end{equation}
therefore,
\begin{equation}\label{s24}
C=\pm\frac{1}{2(1-v^{2})}\left(-k\pm
\sqrt{k^{2}+\frac{4m_{k}(1-v^{2})}{L^{2}}}\right),
\end{equation}
and one can obtain $x^{\prime}$ as a following,
\begin{equation}\label{s25}
x^{\prime}=\frac{vL^{4}}{r^{4}+kL^{2}r^{2}-m_{k}L^{2}}\frac{r_{h}^{4}}{r_{c}^{4}}.
\end{equation}
By integrating it over $r$ and using equation (12) we find $x(r)$
proportional to $\tan^{-1}{r}$. In that case we must use positive
sign of constant $C$ in relation (17) or (24), it shows energy
current from the quark to the horizon. But the negative sign of C
shows energy current from the horizon to the quark where the string
moves  and pulls the quark, so it is a non-physical situation and we
leave this state and give positive sign for the constant $C$.\\
Drag force in this limit can be given by,
\begin{equation}\label{s26}
{\pi}_{x}^{1} = -\frac{T_{0}v}{2L^{2}(1-v^{2})}\left(-k\pm
\sqrt{k^{2}+\frac{4m_{k}(1-v^{2})}{L^{2}}}\right).
\end{equation}
We can neglect the square terms of velocity for heavy quark,
therefore by increasing the speed, drag force increases too.  Then
one can write drag force coefficient as,
\begin{equation}\label{s27}
\mu M = \frac{T_{0}}{2L^{2}}\left(-k\pm
\sqrt{k^{2}+\frac{4m_{k}}{L^{2}}}\right),
\end{equation}
and diffusion constant of quark is,
\begin{equation}\label{s28}
D=\frac{2TL^{2}}{T_{0}m}\left(-k\pm
\sqrt{k^{2}+\frac{4m_{k}}{L^{2}}}\right)^{-1}.
\end{equation}
Here there is another interesting limit to add $q\rightarrow0$ which
is $m_{k}\rightarrow0$. In that case the critical radius obtained by
following relation,
\begin{equation}\label{s29}
r_{c}^{2}=-\frac{kL^{2}}{1-v^{2}},
\end{equation}
in that case we have two values of constant $C$. First is $C=0$
which yields to zero drag force, and second is,
\begin{equation}\label{s30}
C=-\frac{k}{1-v^{2}},
\end{equation}
therefore we have following components of energy and momentum
density,
\begin{eqnarray}\label{s31}
{\pi}_{t}^{1} &=&
-\frac{T_{0}k v^{2}}{L^{2}(1-v^{2})},\nonumber\\
{\pi}_{x}^{1} &=& \frac{T_{0}k v}{L^{2}(1-v^{2})}.
\end{eqnarray}
As before constant $C$ must be positive and on the other hand
critical radius $r_{c}$ should be real, thus in equations (29) and
(30) we must have $v^{2}>1$, so we can't eliminate square term of
velocity. Then we have $\mu M = \frac{T_{0}k }{L^{2}(1-v^{2})}$ and
one can obtain,
\begin{equation}\label{s32}
D=\frac{TL^{2}(v^{2}-1)}{T_{0}k}.
\end{equation}
We will explain both $q\rightarrow0$ and $m_{k}\rightarrow0$ limits
in conclusion.\\
The physical description of loss energy mechanism in terms of some
microscopic parameters in ${\mathcal{N}}=2$ SYM field theory is a
challenge. In $AdS$ dual description energy and momentum current
along the string flow from $D$-brane toward black hole horizon.
Obviously that part of string near horizon has no interaction with
quark surrounding plasma. Therefore energy loss should not be
regarded as a result of scattering in thermal medium. Mentioned
scattering would correspond to small fluctuations in string world
sheet. We will discuss about such small fluctuations in the next
section. In classical string dynamics, there is nothing for
relativistic velocities $v\rightarrow1$, so from equation (32) one
can see $D=0$ for relativistic limit.
\section{Quasi-Normal Modes}
In this section we consider  a  quark moving in the $\mathcal{N}$=2
supergravity thermal plasma without any external field. The goal of
this section is studying behavior of string after long time and with
slow velocity. We consider the dynamics of such a system at a long
time as a small perturbation in the static string which describes
the rest of quark. Indeed quasi-normal modes are classical
perturbations with  non-zero scattering in a given gravitional
background [32-37]. In that case we have only out going boundary
conditions at the black hole horizon which capture the dissipative
nature of the process. Therefore under such a boundary conditions we
have quasi-normal modes of string world sheet. Hence for low
velocity we will neglect $\dot{x}^{2}$ and ${x^{\prime}}^{2}$ in
$-g$, so the equation (5) turns into,
\begin{equation}\label{s33}
\frac{\partial}{\partial r}(\frac{f_{k}
r^{2}}{L^{2}H^{\frac{1}{6}}}x^{\prime})-\frac{H^{\frac{5}{6}}r^{2}}{f_{k}L^{2}}\ddot{x}=0.
\end{equation}
Putting a solution in the form of $x(t,r)=x(r)e^{-\mu t}$,
\begin{equation}\label{s34}
Ox=\mu^{2}x,
\end{equation}
where $O$ is defined as,
\begin{equation}\label{s35}
O=\frac{f_{k}L^{2}}{H^{\frac{5}{6}}r^{2}}\frac{d}{dr}\frac{f_{k}r^{2}}{L^{2}H^{\frac{1}{6}}}\frac{d}{dr}.
\end{equation}
In order to obtain $x(r)$, we expand it in terms of $\mu$ powers,
\begin{equation}\label{s36}
x(r)=x_{0}(r)+\mu x_{1}(r)+\mu^{2} x_{2}(r)+\ldots,
\end{equation}
then we can replace equation (36) in eigenvalue equation (34) to
find,
\begin{eqnarray}\label{s37}
Ox_{0}(r)&=&0,\nonumber\\
Ox_{1}(r)&=&0,\nonumber\\
Ox_{2}(r)&=&x_{0}(r).
\end{eqnarray}
Now boundary conditions can be applied. As we know our solutions
should satisfy Neumann boundary condition so $x^{\prime}(r_{m})=0$.
Thus we select $x_{0}=A$ for the first term in equation (36) where
$A$ is a constant. Then one can apply Neumann boundary condition to
equation (36) and obtain,
\begin{equation}\label{s38}
x^{\prime}(r_{m})=\mu x_{1}^{\prime}(r_{m})+\mu^{2}
x_{2}^{\prime}(r_{m})=0.
\end{equation}
Then by solving equations (37) at the $q\rightarrow0$ and
$m_{k}\rightarrow0$ limits one can find,
\begin{eqnarray}\label{s39}
x_{1}^{\prime}(r)&=&\frac{-AL^{4}}{r^{4}+kL^{2}r^{2}},\nonumber\\
x_{2}^{\prime}(r)&=&\frac{AL^{4}}{r^{4}+kL^{2}r^{2}}\left(r-\sqrt{kL^2}{\tan}^{-1}\frac{r}{\sqrt{kL^{2}}}\right),
\end{eqnarray}
Now by inserting equation (39) into (38), we obtain friction
coefficient $\mu$, and then write diffusion constant as following,
\begin{equation}\label{s40}
D=\frac{T}{M}\left[r_{m}-{\sqrt{kL^{2}}}{\tan}^{-1}\frac{r_{m}}{\sqrt{kL^{2}}}\right].
\end{equation}
In the other word $\mu$ is lowest eigenvalue of operator $O$ or
lowest quasi-normal mode of string which leads us to drag force and
diffusion constant of quark.\\
Now we would like to obtain total energy and momentum of string.
With respect to exponential time-dependent of our solution one can
rewrite equation of motion as following,
\begin{equation}\label{s41}
\frac{\partial}{\partial r}(\frac{f_{k}
r^{2}}{L^{2}H^{\frac{1}{6}}}x^{\prime})=-\mu\frac{H^{\frac{5}{6}}r^{2}}{f_{k}L^{2}}\dot{x}={\mu}^{2}\frac{H^{\frac{5}{6}}r^{2}}{f_{k}L^{2}}x.
\end{equation}
Hence by using equations (6), (8) and (41), we find total momentum
of string,
\begin{equation}\label{s42}
p=\frac{T_{0}}{\mu
L^{4}}\frac{r_{min}^{4}+(q+kL^{2})r_{min}^{2}-m_{k}L^{2}}{(1+\frac{q}{r_{min}^{2}})^{\frac{1}{6}}}x^{\prime}(r_{min}),
\end{equation}
where we use Neumann boundary condition. At $q\rightarrow0$ limit,
we have,
\begin{equation}\label{s43}
p=\frac{T_{0}}{\mu
L^{4}}(r_{min}^{4}+kL^{2}r_{min}^2-m_{k}L^{2})x^{\prime}(r_{min}).
\end{equation}
In order to calculate total energy, we expand $(-g)^{-\frac{1}{2}}$
to second order of velocity and obtain,
\begin{equation}\label{s44}
E=-\frac{T_{0}}{2}\left[\frac{3r_{min}(1+\frac{r_{min}^{2}}{q})^{\frac{1}{6}}}{2(1+\frac{q}{r_{min}^{2}})^{\frac{1}{6}}}
2F_{1}(\frac{2}{3}, \frac{1}{6}, \frac{5}{3},
\frac{-r_{min}^{2}}{q})+\frac{r_{min}^{4}+(q+kL^{2})r_{min}^{2}-m_{k}L^{2}}{L^{4}(1+\frac{q}{r_{min}^{2}})^{\frac{1}{6}}}x(r_{min})x^{\prime}(r_{min})\right],
\end{equation}
where we use Neumann boundary condition too. For the $q\rightarrow0$
limit we have,
\begin{equation}\label{s45}
E=T_{0}\left[r_{m}-r_{min}-
\frac{1}{2L^{4}}(r_{min}^{4}+kL^{2}r_{min}^{2}-m_{k}L^{2})x(r_{min})x^{\prime}(r_{min})\right],
\end{equation}
which is different only in the third term from Ref. [31]. We expect
that relations (43) and (50) together $x(r, t)=x(r) e^{-\mu t}$ lead
us to have simple relation between energy and momentum as
$E=T_{0}(r_{m}-r_{min})+\frac{p^{2}}{2M_{kin}}$, so one can check
validity of this relation easily. We can interpret first term in
right hand side of above equation as $M_{rest}$.
\section{Conclusion}
In this paper, we considered a moving quark in three charges
non-extremal black hole background and obtained drag force by using
$AdS$/CFT correspondence. However,  we set $q_{2}=q_{3}=0$  and
supposed that only non-zero black hole charge is $q_{1}=q$. We
obtained our results for general space with unknown $k$, and here we
would like to represent solutions of different space.\\
First, we consider flat space with $k=0$. In this case the roots of
equation (18) is,
\begin{equation}\label{s46}
{r_{c}}^{2}=-\frac{q}{2}\pm\frac{\sqrt{q^{2}+\frac{4m_{0}L^{2}}{1-v^{2}}}}{2},
\end{equation}
Here, there is an interesting limit as
$\frac{4m_{0}L^{2}}{q^{2}(1-v^{2})}\ll1$. At this limit we have
physical solution for $v^{2}<1$ and one can neglect square terms of
velocity and find,
\begin{equation}\label{s47}
D=\frac{TL^{2}}{T_{0}}(\frac{qL^{2}}{m_{0}^{2}})^{\frac{1}{3}}.
\end{equation}
It means that drag force coefficient $\mu M$ is proportional to
$q^{-\frac{1}{3}}$, for small velocities, similar to equation (26), we find that drag force is proportional to $v$. \\
Second, we consider spherical space i.e, $k=1$. In this
configuration by setting $H\rightarrow H^{3}$,
$m_{k}\rightarrow\eta$, $\Lambda^{2}L^{2}=1$ and
$q=\eta\sinh^{2}{\beta}$ we reduce to Ref. [31], where $\eta$ is
non-extremality parameter, $\Lambda$ is cosmological constant and
$\beta$ is related to electric charge of black hole.\\
we see that $q\rightarrow 0$ limit is corresponding to
$\eta\rightarrow 0$ limit and one can interpret $m_{k}$ as
non-extremality parameter. For this reason, to calculate equations
(39) and (40) we take both limits $q\rightarrow 0$ and
$m_{k}\rightarrow 0$. In another word we may be able to write
$q\propto m_{k}$ (particulary for $k=1$).\\
Third, we consider hyperbolic space corresponding to $k=-1$. This
space is not interesting for us, indeed the only space we can give
some physics is the $S^{5}$ space, because from $AdS/CFT$
correspondence we know that a gauge theory on the boundary of
$AdS_{5}$ space is corresponding to string theory in $AdS_{5}\times
S^{5}$ space. Therefore, it seems the $k=1$ solutions have more
physics in present problem.\\
Here there are several interesting problems in the case of
$q_{1}\neq q_{2}\neq q_{3}$, for example one can obtain drag force
and shear viscosity [28, 38, 39, 40] or jet quenching parameter
[17-25]. Also the effect of higher derivative corrections on drag
force, diffusion constant [41, 42, 43, 44, 45] and the ratio of
shear viscosity $\eta$ to the entropy density $s$ [46, 47, 48] will
be an interesting problem in future.

\end{document}